\documentclass[prd, twocolumn, showpacs, floatfix, letterpaper, nofootinbib, amsmath, amssymb, superscriptaddress]
{revtex4}
\usepackage{graphicx}
\usepackage{epsfig}
\usepackage{bm}
\usepackage{amsfonts}

\usepackage{color}

\newbox\pippobox

\def\be{\begin{equation}}
\def\ee{\end{equation}}
\def\ba{\begin{eqnarray}}
\def\ea{\end{eqnarray}}
\newcommand {\lla} {\ {\raise-.5ex\hbox{$\buildrel<\over\sim$}}\ }
\renewcommand{\(}{\left(}
\renewcommand{\)}{\right)}

\usepackage[T1]{fontenc}
\usepackage[latin1]{inputenc}
\usepackage{graphicx}
\usepackage[english]{babel}
\usepackage{amsmath}
\usepackage{amssymb}
\usepackage{amsfonts}

\def\e{\mathrm{e}}

\def\be{\begin{equation}}
\def\ee{\end{equation}}
\def\bea{\begin{eqnarray}}
\def\eea{\end{eqnarray}}
\newcommand{\ex}{\mathrm{e}}
\newcommand{\dd}{\mathrm{d}}

\def\spose#1{\hbox to 0pt{#1\hss}}

\def\lta{\mathrel{\spose{\lower 3pt\hbox{$\mathchar"218$}}
     \raise 2.0pt\hbox{$\mathchar"13C$}}}
\def\gta{\mathrel{\spose{\lower 3pt\hbox{$\mathchar"218$}}
     \raise 2.0pt\hbox{$\mathchar"13E$}}}

\def\setR{\mathbb{R}}
\def\setC{\mathbb{C}}

\newcommand{\ns}{n_{_\mathrm{S}}}

\newcommand{\GN}{G_{_\mathrm{N}}}

\newcommand{\Mp}{M_{_\mathrm{Pl}}}

\begin{document}

\title{Anisotropy in a Nonsingular Bounce}

\author{Yi-Fu Cai}
\email{yifucai@physics.mcgill.ca}
\affiliation{Department of Physics, McGill University, Montr\'eal, QC H3A 2T8, Canada}

\author{Robert Brandenberger}
\email{rhb@physics.mcgill.ca}
\affiliation{Department of Physics, McGill University, Montr\'eal, QC H3A 2T8, Canada}

\author{Patrick Peter}
\email{peter@iap.fr}
\affiliation{${\cal G}\setR\varepsilon\setC{\cal O}$ -- Institut
d'Astrophysique de Paris, UMR7095 CNRS, Universit\'e Pierre \& Marie Curie,
98 bis boulevard Arago, 75014 Paris, France}


\begin{abstract}
Following recent claims relative to the question of large anisotropy production in regular bouncing scenarios, we study the evolution of such anisotropies in a model where an Ekpyrotic phase of contraction is followed by domination of a Galileon-type Lagrangian which generates a non-singular bounce. We show that the anisotropies decrease during the phase of Ekpyrotic contraction (as expected) and that they can be constrained to remain small during the non-singular bounce phase (a non-trivial result). Specifically, we derive the e-folding number of the phase of Ekpyrotic contraction which leads to a present-day anisotropy in agreement with current observational bounds.
\end{abstract}
\date{\today}
\maketitle

\section{Introduction}

The idea of a non-singular bouncing universe as an alternative to the singular Big Bang paradigm has been attractive on philosophical grounds for a long time. More recently \cite{Wands, Fabio} it has been realized that a non-singular bouncing cosmology with a matter-dominated initial phase of contraction - a ``matter bounce'' cosmology - may provide an alternative to the inflationary paradigm of cosmological structure formation (see e.g. \cite{RHBrev} for recent reviews). Since the curvature perturbation $\zeta$ grows on super-Hubble scales in the contracting phase, fluctuations on long wavelengths get boosted by a larger factor than those on small scales. In a matter-dominated phase of contraction the relative enhancement factor is precisely such as to transform an initial vacuum spectrum into a scale-invariant one. Studies in many non-singular models \cite{models} have shown that the scale-invariance of the spectrum of $\zeta$\footnote{The variable $\zeta$ is the curvature fluctuation in comoving gauge. It is proportional to the canonical fluctuation variable $v$ - the Sasaki-Mukhanov variable \cite{Sasaki, Mukh} - in terms of which the action for cosmological fluctuations has canonical form. See e.g. \cite{MFB} for a comprehensive review and \cite{RHBfluct,PPJPU} for introductory overviews of the theory of cosmological perturbations.} before the bounce is maintained through the bounce on length scales which are larger than the duration of the bouncing phase (the phase in which the violation of the usual energy conditions is localized)\footnote{Note that there are counterexamples as discussed in \cite{DV} and more recently in \cite{BKPPT}.}. Hence, the matter bounce scenario leads to the same shape of the spectrum of primordial curvature fluctuations on super-Hubble scales as inflationary cosmology. Starting with vacuum initial conditions, the fluctuations are also Gaussian and coherent. A specific prediction of the matter bounce with which the scenario can be differentiated from inflation is the shape of the bispectrum \cite{bispectrum}.

Bouncing cosmologies, however, face several problems. First of all, new physics is required to provide the bounce (a pure GR bounce sourced by a regular fluid also leads to dynamical instabilities \cite{PPNPN1}, while with a scalar field a positive spatial curvature \cite{Kpos} or non-canonical kinetic term \cite{Kbounce} is required). Such new physics can either come from introducing new forms of matter such as phantom or quintom fields \cite{quintombounce}, ghost condensates \cite{PPNPN2, Lin},
Galileons \cite{Damien}, S-branes \cite{Kostas} or effective string theory actions \cite{Julio}. It can also arise from corrections to Einstein gravity, as e.g. in the non-singular universe construction of \cite{BMS}, specifically applied to the bouncing scenario in \cite{ModGravBounce}, in the ghost-free higher derivative gravity model of \cite{Tirtho}, in terms of torsion gravity \cite{Cai:2011tc}, or in Ho\v{r}ava-Lifshitz gravity \cite{HLbounce}. Finally, a bouncing phase can also originate from pure quantum cosmological effects \cite{BNS}; those also provide a natural framework in which a dust-dominated contraction is easily implemented to yield a scale-invariant spectrum of perturbation \cite{PertQCdBB}.

A more serious problem for bouncing cosmologies is the fact that the contracting phase is unstable to various effects. As discussed e.g. in \cite{Johanna}, many matter bounce models are unstable to the addition of radiation. This arises since the energy density of radiation scales as $a^{-4}$, where $a(t)$ is the cosmological scale factor, whereas the density of matter grows as $a^{-3}$. In some models (e.g. \cite{quintombounce}) this will lead to a ``Big Crunch'' singularity instead of a smooth bounce. Some constructions (e.g. those of \cite{Lin, BMS}, \cite{Kostas} and \cite{QGBI}) are free from this problem. However, a problem for all constructions mentioned so far is the instability to the growth of anisotropic stress, whose associated energy density grows as $a^{-6}$. This is the famous BKL instability of a contracting universe \cite{BKL}.

A solution of this anisotropy problem can be realized in the Ekpyrotic scenario \cite{Ekp}, in which it is postulated that a matter field with equation of state parameter $w\gg 1$ (where $w = p / \rho$ is the ratio of pressure $p$ to energy density $\rho$) is dominant in the contracting phase\footnote{There are other approaches to address the anisotropy problem. For example, nonlinear matter terms may smooth out the anisotropies \cite{Bozza}. Adding quadratic $R_{\alpha \beta}R^{\alpha \beta}$ terms to the gravitational action can also prevent the BKL instability \cite{Barrow1}.}. Such an equation of state can be realized by treating the dominant form of matter as a scalar field with negative exponential potential. Since the energy density of the dominant matter then scales with $a^{-q}$ with $q \gg 6$, anisotropies become negligible and the BKL instability is avoided 
\cite{EGST}\footnote{Note, however, that including anisotropic pressures may reintroduce instabilities towards anisotropy generation \cite{Barrow2}.}. In a recent paper \cite{us}, a subset of the present authors introduced a scalar field with an Ekpyrotic potential to construct a matter bounce scenario which is free from the BKL instability problem.

The Ekpyrotic scenario in its original formulation \cite{Ekp} involves a singular bounce. In addition, the curvature spectrum of $\zeta$ is an $\ns = 3$ spectrum rather than a scale-invariant $\ns=1$ one \cite{Lyth, Fabio2, Hwang, Peter}. Hence, without non-trivial matching of $\zeta$ across the bounce, one cannot obtain a scale-invariant spectrum at late time\footnote{However, the spectrum of the Bardeen potential $\Phi$ is scale-invariant \cite{Ekp2}, and, as argued in \cite{DV} and shown explicitly in some examples \cite{Tolley, Thorsten}, it is this spectrum which may pass through the bounce, thus yielding a scale-invariant spectrum of curvature fluctuations at late times.}. To solve this problem, a new and non-singular version of the Ekpyrotic scenario \cite{NewEkp} was proposed in which a second scalar field is introduced which does not influence the background dynamics but develops a scale-invariant spectrum which starts out as an isocurvature mode but which is transferred to the adiabatic mode during the evolution. The second field can also be given a ``ghost condensate'' Lagrangian \cite{ghostcond} in which case it mediates a non-singular bounce. However, as has been pointed out in \cite{Stein}, in this ``New Ekpyrotic'' scenario the anisotropies which are highly suppressed during the contracting phase again raise their head and lead to a BKL instability.

In our previous work \cite{us}, we argued qualitatively that in the model we considered the anisotropies remained negligibly small during the bouncing phase. The reason for the difference compared to what happens in the model of \cite{NewEkp} is that in our model the kinetic condensate which grows as the bounce is approached does not need to decrease again by the time of the bounce point. This leads to a shorter bounce time scale and to different dynamics.

In this paper we carefully study the development of anisotropies in the bouncing cosmology with an Ekpyrotic phase of contraction introduced in \cite{us}. We work in the context of a homogeneous but anisotropic Bianchi cosmology in which the scale factors in each spatial dimension evolve independently. We are able to show that no BKL type instability develops, in agreement with what the study of \cite{us} indicated. Our work thus shows that the arguments against non-singular (as opposed to singular) bouncing cosmologies put forwards in \cite{Stein} do not apply to all non-singular bouncing cosmologies.

The outline of this paper is as follows. In the next section we review the bounce model introduced in \cite{us} and derive the resulting equations of motion for a homogeneous but anisotropic universe. In Section 3 we analytically study the background dynamics in each phase of the cosmological evolution from the initial matter phase of contraction through the Ekpyrotic phase to the bouncing phase and the subsequent fast-roll expanding period. Specifically, we determine the decay or growth rates
of the anisotropy parameter in each phase. In Section 4 we solve the dynamical system numerically and present our final results. We close with a general discussion.

A word on notation: We define the reduced Planck mass by $\Mp  = 1/\sqrt{8\pi \GN}$ where $\GN$ is Newton's gravitational constant. The sign of the metric is taken to be $(+,-,-,-)$. Note that we take the value of the mean scale factor at the bounce point to be $a_{_\mathrm{B}} = 1$ throughout the paper.

\section{A nonsingular bounce model}

We consider a nonsingular bounce model in which the universe is filled with two matter components, a cosmic scalar field $\phi$ and a generic matter fluid, as proposed in Ref. \cite{us} (which, in turn, is based on the theory developed in \cite{Vikman}). The Lagrangian of $\phi$ is given by
\begin{eqnarray}\label{Lagrangian}
 \mathcal{L}\left[ \phi \left(x\right)\right] = K(\phi, X) + G(\phi, X) \Box \phi ,
\end{eqnarray}
where $K$ and $G$ are functions of $\phi$ and its canonical kinetic term
\begin{equation}
 X \equiv \frac12 \partial_\mu \phi \partial^\mu \phi ,
\end{equation}
while the other kinetic terms of $\phi$ include the operator
\begin{equation}
 \Box\phi \equiv g^{\mu\nu} \nabla_\mu \nabla_\nu \phi .
\end{equation}

Variation of the above scalar field Lagrangian minimally coupled to Einstein gravity leads to the following corresponding energy momentum tensor
\begin{eqnarray}\label{energystress}
 T^{\phi}_{\mu\nu} &=& (-K+2XG_{,\phi}+G_{,X}\nabla_\sigma{X}\nabla^\sigma\phi)g_{\mu\nu} \nonumber\\
 && + (K_{,X}+G_{,X}\Box\phi-2G_{,\phi})\nabla_\mu\phi\nabla_\nu\phi \nonumber\\
 && - G_{,X}(\nabla_\mu{X}\nabla_\nu\phi+\nabla_\nu{X}\nabla_\mu\phi),
\end{eqnarray}
in which we use the notation that $\mathcal{F}_{,\phi}$ and $\mathcal{F}_{,X}$ denote derivatives of whatever functional $\mathcal{F}(\phi,X)$ may be with respect to $\phi$ and $X$, respectively.

For the model under consideration we choose:
\begin{eqnarray}\label{Kessence}
 K(\phi, X) = \Mp ^2 \left[1-g(\phi) \right]X + \beta X^2 - V(\phi),
\end{eqnarray}
where we introduce a positive-definite parameter $\beta$ so that the kinetic term is bounded from below at high energy scales. Note that the first term of $K$ involves $\Mp ^2$ since in the present paper we adopt the convention that the scalar field $\phi$ is dimensionless.

The function $g(\phi)$ is chosen such that a phase of ghost condensation only occurs during a short time when $\phi$ approaches $\phi=0$. This requires the dimensionless function $g$ to be smaller than unity when $|\phi| \gg 1$ but larger than unity when $\phi$ approaches the origin. To obtain a nonsingular bounce, we must make an explicit choice of $g$ as a function of $\phi$. We want $g$ to be negligible when $|\phi|\gg1$. In order to obtain a violation of the Null Energy Condition after the termination of the Ekpyrotic contracting phase, $g$ must become the dominant coefficient in the quadratic kinetic term when $\phi$ approaches $0$. Thus, we suggest its form to be
\begin{eqnarray}\label{gphi}
 g(\phi) = \frac{2g_0}{\ex^{-\sqrt{\frac{2}{p}}\phi}+\ex^{b_g\sqrt{\frac{2}{p}}\phi}},
\end{eqnarray}
where $g_0$ is a positive constant defined as the value of $g$ at the moment when $\phi=0$, and is required to be larger than unity, $g_0>1$.

\begin{figure}
\includegraphics[width=0.4\textwidth]{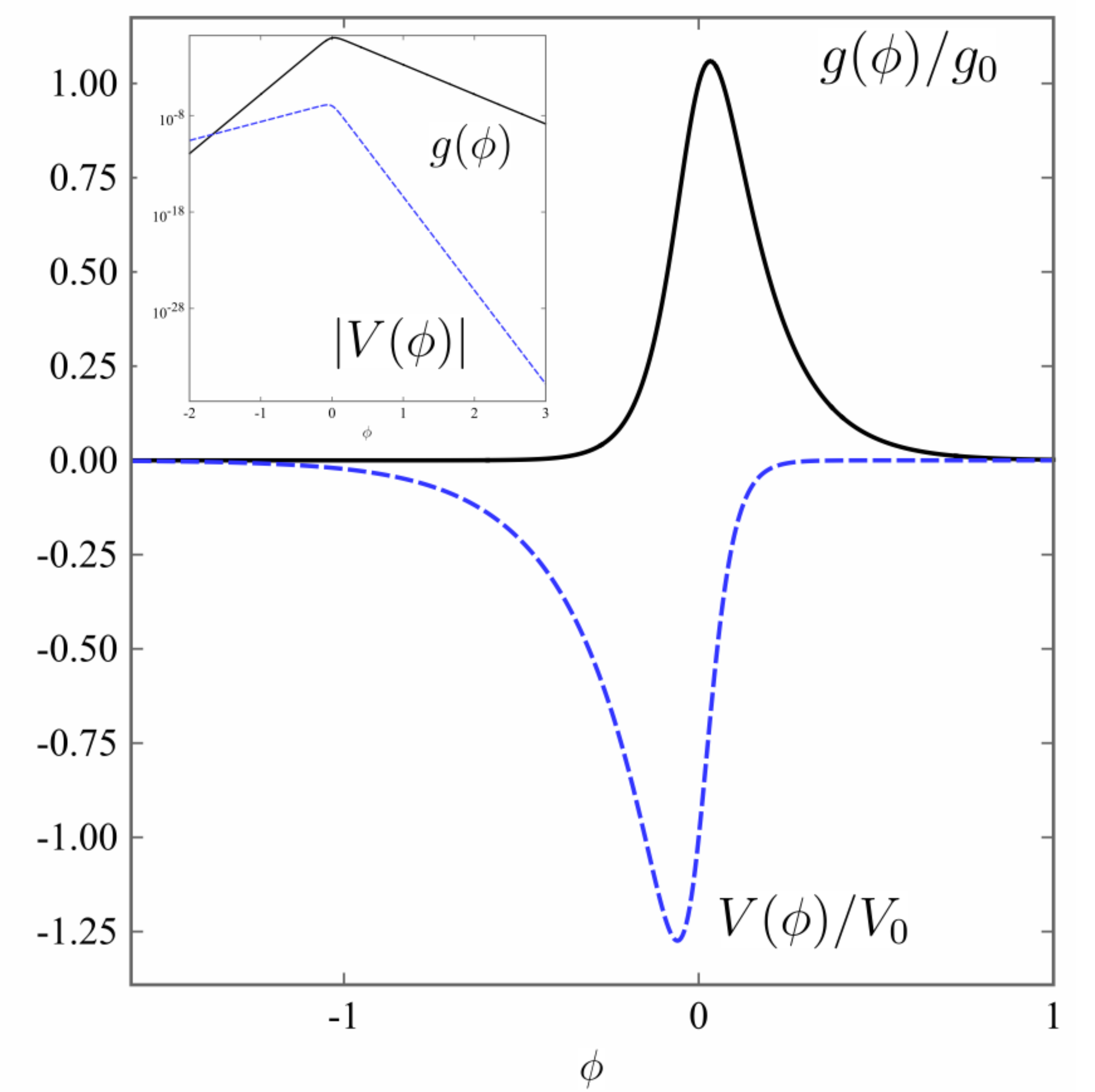}
\caption{Model functions $g(\phi)$ and $V(\phi)$ as given by Eqs.~\eqref{gphi} and \eqref{Vphi}, with background parameters taken as for the following evolution figures, namely as in Eqs. \eqref{parameters1} and \eqref{parameters2}.}
\label{Fig:gVphi}
\end{figure}

We have also introduced a non-trivial potential $V$ for $\phi$. This potential is chosen such that Ekpyrotic contraction is possible. It is well known that the homogeneous trajectory of a scalar field can be an attractor solution when its potential is an exponential function. One example is inflationary expansion of the universe in a positive-valued exponential potential, and the other one is the Ekpyrotic model in which the homogeneous field trajectory for a negative exponential potential is an attractor in a contracting universe. For a phase of Ekpyrotic contraction, we take the form of the potential to be
\begin{eqnarray}\label{Vphi}
 V(\phi) = -\frac{2V_0}{\ex^{-\sqrt{\frac{2}{q}}\phi}+\ex^{b_V\sqrt{\frac{2}{q}}\phi}},
\end{eqnarray}
where $V_0$ is a positive constant with dimension of $({\rm mass})^4$. Thus the potential is always negative and asymptotically approaches zero when $|\phi| \gg1$. Ignoring the second term of the denominator, this potential reduces to the form used in the Ekpyrotic scenario \cite{Ekp}. Both functions $g(\phi)$ and $V(\phi)$ are shown on Fig.~\ref{Fig:gVphi} with the parameters used in the later parts of this work.

The term $G(\phi, X)$ is a Galileon type\footnote{See \cite{Galileonref} for a discussion of Galileon type Lagrangians.} operator which is consistent with the fact that the Lagrangian contains higher order derivative terms in $\phi$, but the equation of motion remains a second order differential equation. Phenomenologically, there are few requirements on the explicit form of $G(\phi, X)$. We introduce this operator since we expect that it can be used to stabilize the gradient term of cosmological perturbations, which requires that the sound speed parameter behaves smoothly and is positive-definite throughout most of the background evolution. For simplicity, we will choose $G$ to be a simple function of only $X$:
\begin{eqnarray}\label{Galileon}
 G(X) = \gamma X,
\end{eqnarray}
where $\gamma$ is a positive-definite number.

We now turn to the study of the cosmology of this model. In order to characterize a homogeneous but anisotropic universe, we take the metric to be of the form
\begin{eqnarray}\label{Bianchimetric}
 \dd s^2 = \dd t^2 - a^2(t) \sum_{i}\ex^{2\theta_i(t)} \sigma^i\sigma^i,
\end{eqnarray}
where $t$ is cosmic time, $\sigma^i$ are linearly independent at all points in space-time and form a three dimensional homogeneous space.

In the case of a Ricci flat space, one can consider the projection $\sigma^i=\dd x^i$ and thus the metric is of Bianchi type-I form. The factor $a(t)$ can be viewed as the mean scale factor of this universe, and the functions $\ex^{\theta_i(t)}$ describe the correction of anisotropies to the scale factor. Since the values of scale factors can be re-scaled arbitrarily, one can impose an additional constraint
\begin{eqnarray}\label{constraint}
 \sum_i \theta_i = 0.
\end{eqnarray}
Then, one can immediately define a mean Hubble parameter as follows,
\begin{eqnarray}
 H \equiv \frac{\dot{a}}{a},
\end{eqnarray}
and the individual Hubble parameters along spatial directions are given by,
\begin{equation}
 H_i\equiv \frac{1}{a\e^{\theta_i}} \frac{\dd}{\dd t} \( a\e^{\theta_i} \)
 = H +\dot\theta_i, \ \ \ \hbox{(no sum)}
\end{equation}
where the overdot denotes the derivative with respect to cosmic time $t$.

Since we are interested in studying anisotropies rather than inhomogeneities we can treat the matter fields to be homogeneous, which implies $\phi$ is only a function of cosmic time. Thus, the kinetic terms of the homogeneous scalar field background become
\begin{eqnarray}
 X &=& \frac12\dot\phi^2, \nonumber \\
 \Box\phi &=& \ddot\phi + 3H\dot\phi ,
\end{eqnarray}
so that, for this background, the energy density of the scalar field is
\begin{eqnarray}\label{rho}
 \rho_\phi = \frac{1}{2}\Mp ^2 (1-g)\dot\phi^2 +\frac{3}{4}\beta\dot\phi^4 +3\gamma H\dot\phi^3 +V(\phi),
\end{eqnarray}
and the pressure is
\begin{eqnarray}\label{pressure}
 p_\phi = \frac{1}{2}\Mp ^2 (1-g)\dot\phi^2 +\frac{1}{4}\beta\dot\phi^4 -\gamma\dot\phi^2\ddot\phi -V(\phi),
\end{eqnarray}
as follows by computing the diagonal components of the stress-energy tensor \eqref{energystress}.

Additionally, the matter fluid contributes its own energy density $\rho_\mathrm{m}$ and pressure $p_\mathrm{m}$, and usually they are associated with a constant equation-of-state parameter $w_\mathrm{m}=p_\mathrm{m}/\rho_\mathrm{m}$. Namely, for normal radiation, $w_\mathrm{m}=\frac13$, while for normal matter, $w_\mathrm{m}=0$.

To derive the equation of motion for $\phi$, one can either vary the Lagrangian with respect to $\phi$ or, equivalently, require that the covariant derivative of its stress-energy tensor vanishes. This yields
\begin{eqnarray}\label{eom}
 {\cal P} \ddot\phi + {\cal D} \dot\phi +V_{,\phi} = 0,
\end{eqnarray}
where we have introduced
\begin{eqnarray}
\label{Pterm}
 {\cal P} &=& (1-g)\Mp ^2 +6\gamma H\dot\phi +3\beta\dot\phi^2 +\frac{3\gamma^2}{2\Mp ^2}\dot\phi^4,\\
\label{Fterm}
 {\cal D} &=& 3(1-g)\Mp ^2H +\left(
 9\gamma{H}^2-\frac{1}{2}\Mp ^2g_{,\phi}\right)
 \dot\phi +3\beta{H}\dot\phi^2 \nonumber\\
 && -\frac{3}{2}(1-g)\gamma\dot\phi^3 -\frac{9\gamma^2H\dot\phi^4}{2\Mp ^2} -\frac{3\beta\gamma\dot\phi^5}{2\Mp ^2} \nonumber\\
 && -\frac{3}{2}G_{,X}\sum_i\dot\theta_i^2\dot\phi - \frac{3G_{,X}}{2\Mp ^2}(\rho_\mathrm{m}+p_\mathrm{m})\dot\phi .
\end{eqnarray}
{}From Eq.~\eqref{eom}, it is clear that the function ${\cal P}$ determines the positivity of the kinetic term of the scalar field and thus can be used to determine whether the model contains a ghost or not at the perturbative level; the function ${\cal D}$ on the other hand, represents an effective damping term. By keeping the first terms of the expressions of
${\cal P}$ and ${\cal D}$ and setting $g = 0$, one can recover the standard Klein-Gordon equation in the FRW background. Neglecting the other terms is a good approximation when the velocity of $\phi$ is sub-Planckian. Note that the friction term ${\cal D}$ contains the contributions from anisotropic factors and matter fluid, which can be suppressed for small values of
$\dot\phi$. However, these terms will become important during the bouncing phase where $\dot\phi$ reaches a maximal value. For simplicity, in the following we will consider matter fluid is cold and thus $w_\mathrm{m}=0$.

{F}inally, we can write down Einstein equations in this background, given by
\begin{eqnarray}
 \Mp ^2 \left(R_{\mu\nu}-\frac{R}{2}g_{\mu\nu} \right) = T^\phi_{\mu\nu} +T^m_{\mu\nu} .
\end{eqnarray}
Once expanded in components, this tensor equation yields the effective Friedmann equations,
\begin{eqnarray}
\label{Friedmann1} H^2 &=& \frac{\rho_{_\mathrm{T}}}{3\Mp ^2} + \frac{1}{6}\sum_i\dot\theta_i^2,\\
\label{Friedmann2} \dot{H} &=& -\frac{\rho_{_\mathrm{T}} + p_{_\mathrm{T}}}{2\Mp ^2} - \frac{1}{2}\sum_i\dot\theta_i^2,
\end{eqnarray}
where $\rho_{_\mathrm{T}}$ and $p_{_\mathrm{T}}$ represent the total energy density and pressure in the Bianchi type-I universe, i.e., the sum of the contributions of the scalar field and the fluid.

Moreover, combining the spatial component of Einstein equation with the constraint equation \eqref{constraint} yields
\begin{eqnarray}\label{eomani}
 \ddot\theta_i + 3H\dot\theta_i = 0,
\end{eqnarray}
{}from which it follows that
\begin{eqnarray}\label{dotthetai}
 \dot\theta_i(t) = M_{\theta,i}\frac{a_{_\mathrm{B}}^3}{a^3(t)},
\end{eqnarray}
where $a_{_\mathrm{B}}$ is the mean scale factor of the universe at the bouncing point. The coefficients $M_{\theta,i}$ are integral constants with a dimension of mass. According to the constraint equation \eqref{constraint}, one can read off that
\begin{eqnarray}
 \sum_i M_{\theta,i}=0.
\end{eqnarray}

Plugging Eq.~\eqref{dotthetai} into Eq. \eqref{Friedmann1} shows that one can introduce an effective energy density of anisotropy
\begin{eqnarray}
 \rho_\theta \equiv \frac{\Mp ^2}{2}\sum_i\dot\theta_i^2 \propto a^{-6},
\end{eqnarray}
whose evolution as $1/a^6$ implies an effective equation-of-state parameter equal to $w_\theta=1$. We see that this effective energy density increases faster than that of pressureless matter or radiation in a contraction universe. This is the source of the BKL instability of the contracting phase of many bouncing cosmologies.

\section{Background evolution}

The initial conditions of our model are chosen (as in \cite{us}) such that we start in a contracting phase dominated by regular matter. Since the energy density of the Ekpyrotic scalar field $\phi$ grows faster than that of regular matter, $\phi$ will at some time begin to dominate the energy density. At this point, the Ekpyrotic phase of contraction begins, and lasts until the nonsingular bounce interval begins (this is the phase where the new physics effects dominate), followed by a period of fast-roll expansion, which in turn ends at a transition to the expansion of Standard Big Bang cosmology. We choose the initial conditions for the density of regular matter and for the value of $\phi$ such that the temperature at which the Ekpyrotic phase begins is higher than that at the time of equal matter and radiation in the Standard Big Bang expanding phase. In this way, we ensure that initial vacuum perturbations develop into a scale-invariant spectrum of cosmological fluctuations on all scales which are currently probed.

In the following we study the evolution of the anisotropy in each of the periods of cosmological evolution mentioned above.

\subsection{Matter contraction}

We start by considering the period when the universe is dominated by a pressureless matter fluid, i.e., when the background equation of state parameter is roughly $w=0$. In this phase, the mean scale factor evolves as
\begin{eqnarray}
 a(t) \simeq a_{_\mathrm{E}} \left( \frac{t-\tilde{t}_{_\mathrm{E}}}{t_{_\mathrm{E}} -\tilde{t}_{_\mathrm{E}}} \right)^{2/3},
\end{eqnarray}
where $t_{_\mathrm{E}}$ denotes the final moment of matter contraction and the beginning of the Ekpyrotic phase, and $a_{_\mathrm{E}}$ is the value of the mean scale factor at the time $t_{_\mathrm{E}}$. In the above, $\tilde{t}_{_\mathrm{E}}$ is an integration constant which is introduced to match the mean Hubble parameter continuously at the time $t_{_\mathrm{E}}$, i.e.,
\begin{eqnarray}
 \tilde{t}_{_\mathrm{E}} \simeq t_{_\mathrm{E}}-\frac{2}{3H_{_\mathrm{E}}}.
\end{eqnarray}
Correspondingly, the mean Hubble parameter during the period of matter contraction is given by
\begin{eqnarray}\label{Hubble_c}
 H(t) = \frac{2}{3(t-\tilde{t}_{_\mathrm{E}})}.
\end{eqnarray}

Following Eq. \eqref{dotthetai}, one can immediately write down the time derivatives of the anisotropy factors as follows
\begin{eqnarray}\label{dotthetai_m}
 \dot\theta_i(t) = M_{\theta,i} \frac{a_{_\mathrm{B}}^3 H^2(t)}{a_{_\mathrm{E}}^3 H_{_\mathrm{E}}^2},
\end{eqnarray}
where $H_{_\mathrm{E}}$ is the value of mean Hubble parameter at the moment $t_{_\mathrm{E}}$. Note that the dimensional parameters of anisotropy $M_{\theta,i}$ are therefore related to the values of $\dot\theta_i$ at the moment $t_{_\mathrm{E}}$; a smooth bounce will thus take place only if these values satisfy some constraints which we derive below. 

Integrating the above yields the following expressions for the anisotropy factors
\begin{equation}
\label{thetai_m}
 \theta_i = \int_{-\infty}^t \!\!\!\!\!\!\dot\theta_i(t')\dd t'
 = -\frac{2a_{_\mathrm{B}}^3M_{\theta,i}}{3a_{_\mathrm{E}}^3H_{_\mathrm{E}}}
 \frac{t_{_\mathrm{E}}-\tilde{t}_{_\mathrm{E}}}{t-\tilde{t}_{_\mathrm{E}}}
 = -\frac{2a_{_\mathrm{B}}^3M_{\theta,i}}{3a_{_\mathrm{E}}^3H_{_\mathrm{E}}^2}H(t).
\end{equation}
{}From Eq. \eqref{thetai_m}, we can observe that the anisotropy factors $\theta_i$ approach zero in the limit of $ t \rightarrow -\infty $. As the universe contracts with a matter-dominated equation of state, the absolute value of $H$ increases and thus $\theta_i$ becomes larger as well.

Before the Ekpyrotic phase, the universe is dominated by the pressureless matter fluid and the energy density of normal matter fluid evolves as $\rho_\mathrm{m} \simeq 3\Mp ^2 H^2$. However, Eq.~\eqref{dotthetai_m} shows that the effective energy density of anisotropies evolves as $\rho_\theta \sim H^4$ and will therefore come to dominate. This is nothing but another way of stating the BKL instability in a contracting matter-dominated universe. In the following we will see that the Ekpyrotic phase of contraction cures this problem. However, in order to be able to enter this phase, the initial anisotropy cannot be too large for the model not to break down altogether. The requirement is that $\rho_\theta$ is less than $\rho_\mathrm{m}$ at the transition time $t_{_\mathrm{E}}$, which implies
\begin{eqnarray}\label{constraint_1}
 \sum_i M_{\theta,i}^2 \lesssim 6 H_{_\mathrm{E}}^2 \frac{a_{_\mathrm{E}}^6}{a_{_\mathrm{B}}^6} ~,
\end{eqnarray}
a requirement which we implement in our numerical discussion below.

Let us now move on to the discussion of the evolution of the anisotropy factors during the Ekpyrotic phase of contraction.

\subsection{Ekpyrotic contraction}

We assume a homogeneous scalar field $\phi$ which is initially placed in the regime of $\phi \ll -1$ in the phase of matter contraction. In this case, the Lagrangian for $\phi$ approaches the conventional canonical form and thus yields an attractor solution which is given by 
\begin{eqnarray}
 \phi(t) \simeq -\sqrt{\frac{q}{2}} \ln \left[ \frac{2V_0(t-\tilde{t}_{_{\mathrm{B}-}})^2}{q(1-3q)\Mp ^2} \right],
\end{eqnarray}
where $\tilde{t}_{_{\mathrm{B}-}}$ is an integration constant which is chosen in order that the mean Hubble parameter at the end of the phase of Ekpyrotic contraction matches with the one at the beginning of the bouncing phase. This attractor solution corresponds to an effective equation of state
\begin{eqnarray}
 w \simeq -1+\frac{2}{3q}.
\end{eqnarray}

During the phase of Ekpyrotic contraction, the mean scale factor evolves as
\begin{eqnarray}
 a(t) \simeq a_{_{\mathrm{B}-}} \left(\frac{t-\tilde{t}_{_{\mathrm{B}-}}}{t_{_{\mathrm{B}-}}
 -\tilde{t}_{_{\mathrm{B}-}}}\right)^q,
\end{eqnarray}
where $a_{_{\mathrm{B}-}}$ is the value of mean scale factor at the time $t_{_{\mathrm{B}-}}$ which corresponds to the end of Ekpyrotic contraction and the beginning of the bouncing phase. Therefore, the mean Hubble parameter is given by
\begin{eqnarray}\label{Hubble_E}
 H(t) \simeq \frac{q}{t-\tilde{t}_{_{\mathrm{B}-}}},
\end{eqnarray}
where, in order to make $H(t)$ continuous at the time $t_{_{\mathrm{B}-}}$, one must set
\begin{eqnarray}
 \tilde{t}_{_{\mathrm{B}-}} = t_{_{\mathrm{B}-}}-\frac{q}{H_{_{\mathrm{B}-}}}.
\end{eqnarray}
Additionally, we require the mean scale factor to evolve smoothly and continuously at the time $t_{_\mathrm{E}}$. This leads to the relation
\begin{eqnarray}\label{a_E}
 a_{_\mathrm{E}} \simeq a_{_{\mathrm{B}-}} \left(\frac{H_{_{\mathrm{B}-}}}{H_{_\mathrm{E}}}\right)^q.
\end{eqnarray}

Given these preliminaries, we find that the time derivatives of the anisotropy factors evolve as
\begin{eqnarray}
 \dot\theta_i(t) &\simeq& M_{\theta,i} \frac{a_{_\mathrm{B}}^3}{a_{_{\mathrm{B}-}}^3} \left( \frac{H}{H_{_\mathrm{E}}} \right)^{3q},
\end{eqnarray}
and thus the effective energy density of anisotropy evolves as $\rho_\theta \sim H^{6q}$. Hence, whereas $\rho_\theta$ still increases during the phase of Ekpyrotic contraction, the growth rate is much slower than in the matter-dominated phase (given the small value of $q$). In particular, the energy density in $\phi$ increases much faster than that due to the anisotropies: the BKL instability is tamed in this manner by the Ekpyrotic contraction phase.

Integrating the above expressions for $\dot\theta_i$ we obtain
\begin{eqnarray}\label{thetai_E}
 \theta_i(t) &=& \theta_{i, E} +\int_{t_{_\mathrm{E}}}^t\dot\theta_i(t')\dd t' \nonumber\\
 &\simeq& \frac{(2-3q) a_{_\mathrm{B}}^3 M_{\theta,i}}{3(1-3q)a_{_\mathrm{E}}^3H_{_\mathrm{E}}} \left[ -1+ \frac{3q}{2-3q}\left(\frac{H_{_\mathrm{E}}}{H}\right)^{1-3q} \right].\cr & &
\end{eqnarray}
{}From the expression \eqref{thetai_E} we see that the absolute values of $\beta_i$ are monotonically increasing when cosmic time $t$ evolves from $t_{_\mathrm{E}}$ to the onset of the bouncing phase at time $t_{_{\mathrm{B}-}}$. However, they very rapidly converge to their limiting values, and since any constant part of the anisotropy factors can be absorbed in a rescaling of the spatial coordinates, it turns out that the actual anisotropy, effectively measured as the distance to an effective FLRW space, is effectively decreasing as the amplitude of the mean Hubble parameter increases: the Ekpyrotic phase thus provides a simple solution to the BKL instability growth in a contracting phase.

\subsection{Bounce phase}

In our model the scalar field evolves monotonically from $\phi \ll -1$ to $\phi \gg 1$. For values of $\phi$ between $\phi_- \sim -\sqrt{p/2} \ln (2g_0)$ and $\phi_+\sim \sqrt{p/2} \ln (2g_0)/b_g$ (assuming one term in the denominator of $g(\phi)$ dominates over the other at each transition time), the value of the function $g(\phi)$ becomes larger 
than unity and thus the universe enters a ghost condensate state. The occurrence of the ghost condensate naturally yields a short period of null energy condition violation and this in turn gives rise to a nonsingular bounce.

As shown in Ref. \cite{us}, we have two useful parameterizations to describe the evolution of the scale factor in the bounce phase. One is the linear parametrization of the mean Hubble parameter
\begin{eqnarray}
 H(t) \simeq \Upsilon t,
\end{eqnarray}
and the other is the evolution of the background scalar
\begin{eqnarray}\label{dotphi_bouncing}
 \dot\phi(t) \simeq \dot\phi_{_\mathrm{B}} \ex^{-t^2/T^2},
\end{eqnarray}
where the coefficient $\Upsilon$ is set by the detailed microphysics of the bounce. The coefficient $T$ can be determined by matching the detailed evolution of the scalar field at the beginning or the end of the bouncing phase, which will be addressed in next subsection. Thus, during the bounce the mean scale factor evolves as
\begin{eqnarray}
 a(t) \simeq a_{_\mathrm{B}} \ex^{\frac12 \Upsilon t^2}.
\end{eqnarray}

Note that a nonsingular bounce requires that the total energy density vanishes at the bounce point. The total energy density includes the contributions from the matter fields and the anisotropy factors. This leads to the following result for the value of $\dot\phi_{_\mathrm{B}}$
\begin{eqnarray}
 \dot\phi_{_\mathrm{B}}^2 &\simeq& \frac{(g_0-1)\Mp ^2}{3\beta} \left[ 1+\sqrt{1+\frac{12\beta(V_0+\rho_\mathrm{m}+\rho_\theta)}{(g_0-1)^2\Mp ^4}} \,\right] \nonumber\\
 &\simeq& \frac{2(g_0-1)}{3\beta}\Mp ^2,
\end{eqnarray}
where we have made use of approximations that $\rho_\mathrm{m}$ and $\rho_\theta$ are much less than $V_0$ and $V_0\ll \Mp^4$ in the second line. These approximations must be valid for the model to hold since both $\rho_\mathrm{m}$ and $\rho_\theta$ are greatly diluted in Ekpyrotic phase and $V_0$ is the maximal absolute value of the potential of $\phi$ which, according to the observational constraint from the amplitude of cosmological perturbations, must be far below the Planck scale.

Now we calculate the anisotropy factors in the bouncing phase. We first study their time derivatives, which are given by
\begin{eqnarray}
 \dot\theta_i(t) \simeq M_{\theta,i} \ex^{-\frac{3}{2}\Upsilon t^2}.
\end{eqnarray}
Thus we see that the integration constants $M_{\theta,i}$ can be interpreted as the values of $\dot\theta_i$ at the bounce point $t_{_\mathrm{B}} = 0$, and they are the maximal values $\dot\theta_i$ will ever take throughout the whole cosmic evolution. It is again easy to perform the integrals and obtain the anisotropy factors:
\begin{eqnarray}\label{theta_i_bouncing}
 \theta_i(t) \simeq \theta_{i,_{\mathrm{B}-}}+M_{\theta,i}\sqrt{\frac{\pi}{6\Upsilon}}
 ~ {\rm Erf} \left( \sqrt{\frac{3\Upsilon}{2}}t\right) \Bigg|^t_{t_{_{\mathrm{B}-}}},
\end{eqnarray}
where ${\rm Erf}$ is the Gauss error function. Around the bounce point, we can perform a Taylor expansion of \eqref{theta_i_bouncing} up to leading order and thus obtain the following approximate expression,
\begin{eqnarray}
 \theta_i(t) \simeq \theta_{i,_{\mathrm{B}-}}+M_{\theta,i} \left(t-t_{_{\mathrm{B}-}}\right),
\end{eqnarray}
which is a linear function of cosmic time. We find that the growth of the anisotropy at linear order only depends on the coefficients $M_{\theta,i}$. However, at nonlinear order in cosmic time, the growth would also depend on the slope of the bounce phase characterized by the parameter $\Upsilon$.

Finally, we have the relations
\begin{eqnarray}
 a_{_{\mathrm{B}-}} &\simeq& a_{_\mathrm{B}} \ex^{\frac{\Upsilon}{2}t_{_{\mathrm{B}-}}^2},\\
 H_{_{\mathrm{B}-}} &\simeq& \Upsilon t_{_{\mathrm{B}-}},
\end{eqnarray}
by matching the mean scale factor and the mean Hubble parameter at the beginning of the bouncing phase.

\subsection{Fast-roll expanding phase}

After the bounce, the universe enters the expanding phase, which typically begins with a period of fast-roll expansion for the background during which the universe is  still dominated by the scalar field $\phi$. During this stage, the motion of $\phi$
is dominated by its kinetic term while the potential is negligible. Thus, the background equation of state parameter is $w \simeq 1$. Correspondingly, the mean scale factor evolves as
\begin{eqnarray}
 a(t) \simeq a_{_{\mathrm{B}+}} \left(\frac{t-\tilde{t}_{_{\mathrm{B}+}}}
 {t_{_{\mathrm{B}+}}-\tilde{t}_{_{\mathrm{B}+}}}\right)^{1/3},
\end{eqnarray}
where $t_{_{\mathrm{B}+}}$ represents the end of the bounce phase and the beginning of the fast-roll period, and $a_{_{\mathrm{B}+}}$ is the value of the mean scale factor at that moment. Then one can write down the mean Hubble parameter in the fast-roll phase
\begin{eqnarray}
 H(t) \simeq \frac{1}{3(t-\tilde{t}_{_{\mathrm{B}+}})},
\end{eqnarray}
and the continuity of the mean Hubble parameter at $t_{_{\mathrm{B}+}}$ yields
\begin{eqnarray}
 \tilde{t}_{_{\mathrm{B}+}} = t_{_{\mathrm{B}+}} -\frac{1}{3H_{_{\mathrm{B}+}}}.
\end{eqnarray}

Recall that, in Eq. \eqref{dotphi_bouncing}, we made use of a Gaussian parametrization of the scalar field evolution in the bounce phase, with characteristic timescale $T$. In the fast roll phase we find the following approximate solution for the evolution of $\phi$:
\begin{eqnarray}
 \dot\phi(t) \simeq \dot\phi_{_{\mathrm{B}+}} \frac{a_{_{\mathrm{B}+}}^3}{a^3(t)}
 \simeq \dot\phi_{_\mathrm{B}} \ex^{-{t_{_{\mathrm{B}+}}^2}/{T^2}}\frac{H(t)}{H_{_{\mathrm{B}+}}},
\end{eqnarray}
where we have applied \eqref{dotphi_bouncing} in the second equality. This implies that
\begin{eqnarray}
 \rho_\phi \simeq \frac{\Mp ^2}{2}\dot\phi^2 \simeq \frac{\Mp ^2\dot\phi_{_\mathrm{B}}^2}
 {2\ex^{2t_{_{\mathrm{B}+}}^2/T^2}} \frac{H^2}{H_{_{\mathrm{B}+}}^2}.
\end{eqnarray}
Moreover, the Friedmann equation requires that $\rho_\phi\simeq 3\Mp ^2H^2$ in the fast-roll phase, so that $T^2$ is given by
\begin{eqnarray}
 T^2 \simeq \frac{2H_{_{\mathrm{B}+}}^2}{\Upsilon^2\ln\left[\displaystyle
 \frac{\Mp ^2(g_0-1)}{9\beta H_{_{\mathrm{B}+}}^2}\right]}.
\end{eqnarray}

During the fast-roll expansion epoch, one can solve for the time derivatives of the anisotropy factors
\begin{eqnarray}\label{dot_theta_FR}
 \dot\theta_i(t) \simeq M_{\theta,i}\frac{a_{_\mathrm{B}}^3H(t)}{a_{_{\mathrm{B}+}}^3H_{_{\mathrm{B}+}}},
\end{eqnarray}
from which it follows that the effective energy density of anisotropy evolves as $\rho_{\theta} = \rho_{\theta,_{\mathrm{B}+}} H^2/H_{_{\mathrm{B}+}}^2$, and thus its ratio relative to the total energy density does not change (recall that the absolute value of this ratio is also very small because of the decrease of the shear contribution during the Ekpyrotic phase of contraction discussed above). On the other hand, the matter fluid, having $w_\mathrm{m}=0$, sees its energy density going as
$\rho_\mathrm{m} = \rho_{\mathrm{m},_{\mathrm{B}+}} H/H_{_{\mathrm{B}+}}$, and therefore its contribution to the background will catch up with that of the scalar field and the anisotropy, bringing the fast roll phase to an end and initiating the transition to the usual expansion history of Standard Big Bang cosmology.

The anisotropy factors in the fast roll phase are given by
\begin{eqnarray}
 \theta_i(t) \simeq \theta_{i,_{\mathrm{B}+}} + \frac{a_{_\mathrm{B}}^3M_{\theta,i}}{3a_{_{\mathrm{B}+}}^3
 H_{_{\mathrm{B}+}}} \ln\left( \frac{t-\tilde{t}_{_{\mathrm{B}+}}}{t_{_{\mathrm{B}+}}-\tilde{t}_{_{\mathrm{B}+}}}
 \right),
\end{eqnarray}
so they are safely growing only logarithmically during this phase.

To conclude this section, it is clear that the anisotropy contribution can easily be made to remain small throughout the entire evolution, i.e. including the matter contraction and the bounce phases. We now turn to the actual constraints that should be imposed on the parameters of our model.

\subsection{Theoretical constraints}

In this subsection, we study the theoretical constraints on the scenario of anisotropic nonsingular bounce. It is convenient to introduce an effective e-folding number of the Ekpyrotic phase as follows
\begin{eqnarray}\label{N_E}
 {\cal N}_{_\mathrm{E}} \equiv \ln \left( \frac{a_{_{\mathrm{B}-}}
 H_{_{\mathrm{B}-}}}{a_{_\mathrm{E}}H_{_\mathrm{E}}} \right) = (1-q) \ln\left(
 \frac{H_{_{\mathrm{B}-}}}{H_{_\mathrm{E}}}\right),
\end{eqnarray}
which characterizes the variation of the mean value of the conformal Hubble scale $aH$ during this period.

Recall that the anisotropy is not allowed to dominate before the onset of the Ekpyrotic phase. This is the constraint \eqref{constraint_1}. Applying \eqref{a_E} and \eqref{N_E}, we can further derive the following relation for the anisotropy at the bounce time:
\begin{eqnarray}\label{N_E_app}
 \sum_i \frac{M_{\theta,i}^2}{6H_{_{\mathrm{B}-}}^2} \simeq
 \ex^{-2(1-3q)\mathcal{N}_{_\mathrm{E}}/(1-q)},
\end{eqnarray}
which is, as expected, exponentially suppressed by the number of e-foldings of Ekpyrotic contraction. Note that one can introduce a relative density parameter to describe the contribution of the anisotropies as follows
\begin{eqnarray}
 \Omega_\theta \equiv \frac{\rho_\theta}{3\Mp ^2H^2}.
\end{eqnarray}
This parameter must be less than the amplitude of the observed anisotropies in the cosmic microwave background (CMB), which is of order $\Omega_\mathrm{pert} \sim \mathcal{O}(10^{-5})$, hereby defining $\Omega_\mathrm{pert}$ as the amount of energy density in perturbations, as given by CMB observations. This leads to the following constraint on the number of e-foldings
\begin{eqnarray}\label{N_E_con_pre}
 \sum_i \frac{{M_{\theta,i}}^2}{6H_{_{\mathrm{B}-}}^2} < \Omega_\mathrm{pert},
\end{eqnarray}
where we have made use of \eqref{dot_theta_FR}. Provided that this constraint is satisfied, the current anisotropy will be below the current upper bound provided that it was initially small enough not to prevent the onset of Ekpyrotic contraction. Combining this inequality and the approximate relation \eqref{N_E_app}, we can obtain a lower bound on the e-folding number of the Ekpyrotic phase
\begin{eqnarray}\label{constraint_NE}
 {\cal N}_{_\mathrm{E}} > \frac{1-q}{2(1-3q)}\ln\left(\frac{1}{\Omega_\mathrm{pert}}\right).
\end{eqnarray}
Note that the nonsingular bounce model studied in the present paper belongs to ``fast bounce'' category and the absolute values of $H_{_{\mathrm{B}-}}$ and $H_{_{\mathrm{B}-}}$ are of the same order. Thus, we can simply consider $H_{_{\mathrm{B}-}}$ to be the energy scale of the bounce. As an example, if we choose $q=0.1$ and $\Omega_\mathrm{pert} = 10^{-5}$, then the relation \eqref{constraint_NE} yields ${\cal N}_{_\mathrm{E}} > 7.4$. This result implies that the Ekpyrotic phase is rather efficient at lowering the contribution of anisotropies since a mere few e-folds of Ekpyrotic
contraction are enough to damp any reasonable initial anisotropy to negligible level.

Note that the e-folding number ${\cal N}_{_\mathrm{E}}$ can also be constrained on observable scales of CMB experiments, and we expect this may impose a much more stringent constraint on the anisotropy parameters. For that, one could require that the anisotropy contribution be smaller that the observed level of perturbations, namely $\Omega_\mathrm{pert}\sim 10^{-10}$.

It is interesting to note that increasing the Ekpyrotic e-fold number to, say, ${\cal N}_{_\mathrm{E}} \sim 15$ while keeping $q\sim0.1$, leads to relation \eqref{constraint_1} to yield $\Omega_{\theta} \sim 10^{-10} \sim \Omega_\mathrm{pert}^2 \ll \Omega_\mathrm{pert}$. This means that after a sufficiently long Ekpyrotic contraction, the anisotropy contribution is totally negligible, even compared with the amplitude of primordial perturbations. At this particular level however, the anisotropy contribution could be comparable to second order in terms of the primordial curvature perturbation expansion, and thus could contribute to the primordial non-Gaussianities of local shape with $f_{_\mathrm{NL}}\sim \mathcal{O}(1)$.

If one wants to make successful contact with late time cosmology, there is a second constraint that should be implemented on the model, namely that the fast roll phase ends before the time of Big Bang Nucleosynthesis (BBN). Roughly speaking, the energy density of regular matter does not change much during the phase of Ekpyrotic contraction (for small values of $q$). On the other hand, the density of $\phi$ grows rapidly. In the fast roll phase of expansion, the decrease in the density of regular matter is no longer negligible. Hence, the energy density of matter will be much lower at the time $t_{_\mathrm{F}}$ when $\rho_\mathrm{m}(t_{_\mathrm{F}}) = \rho_{\phi}(t_{_\mathrm{F}})$ than at the time $t_{_\mathrm{E}}$ when Ekpyrotic contraction begins. In fact, it is straightforward to derive that
\begin{eqnarray}\label{constraint_2}
 H_{_\mathrm{F}} \lesssim |H_{_\mathrm{E}}|\ex^{-(1-3q)\mathcal{N}_{_\mathrm{E}}/(1-q)},
\end{eqnarray}
showing that the value of $H_{_\mathrm{F}}$ should be much less than $|H_{_\mathrm{E}}|$. The Hubble rate $H_{_\mathrm{F}}$ is associated with the initial temperature $T_{_\mathrm{F}}$ when the expansion begins to follow the Standard Big Bang evolution (it is the equivalent of the temperature of reheating in inflationary cosmology). Specifically, the relation is
\begin{equation}
H_{_\mathrm{F}} \simeq  \frac{g_s^{1/2} \pi T_{_\mathrm{F}}^2}{9.5\Mp},
\end{equation}
where $g_s$ is the effective partible number for radiation. As a consequence, in analogy with inflationary cosmology, the constraint \eqref{constraint_2} leads to an upper bound on the effective ``reheating'' temperature:
\begin{eqnarray}\label{constraint_T}
 T_{_\mathrm{F}} \lesssim \left(\frac{3\Mp |H_{_{\mathrm{B}-}}|}{g_s^{1/2}}\right)^{\frac{1}{2}}
\ex^{-(2-3q)\mathcal{N}_{_\mathrm{E}}/[2(1-q)]}
\end{eqnarray}
in our nonsingular bounce model. From the BBN constraint, we find that the lower limit of the ``reheating'' temperature is of the order $\mathcal{O}({\rm MeV})$. If we consider this lower bound and take $g_s\sim100$, ${\cal N}_{_\mathrm{E}}\sim30$ and
$q\sim0.1$, then we find that $H_{_{\mathrm{B}+}} > 10^{-17}\Mp $ which can easily be implemented in the model, as we shall see in the following numerical calculations.

\subsection{Numerical estimates}

To illustrate that a nonsingular bounce can be achieved in our model, we numerically solved the background equations of motion. Expressing all relevant functions and parameters in the corresponding units of the reduced Planck mass $\Mp$, we set
\begin{eqnarray}\label{parameters1}
 &V_0 = 10^{-7},~~g_0 = 1.1,~~\beta = 5,~~\gamma = 10^{-3},\nonumber\\
 &b_V = 5,~~b_g = 0.5,~~p = 0.01,~~q=0.1
\end{eqnarray}
to illustrate the calculations.

\begin{figure}
\includegraphics[scale=0.3]{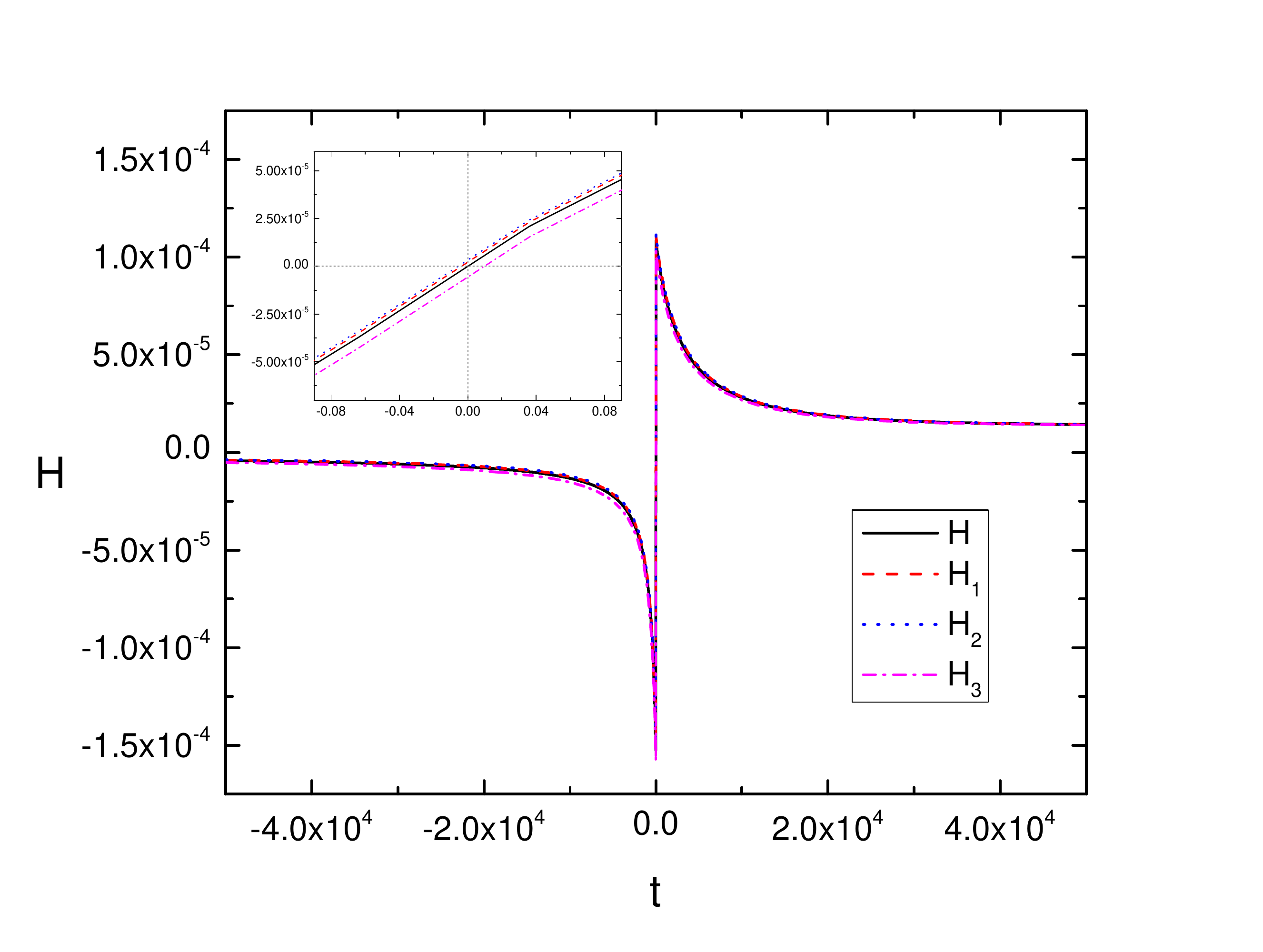}
\caption{Time evolution of the Hubble parameters $H$ (black line) and $H_i$ (red dashed, blue dotted and magenta dot-dashed lines for the Hubble expansion rates along the $x_1$, $x_2$, and $x_3$ axes, respectively), in units of the reduced Planck mass $\Mp$, with background parameters given by Eqs. \eqref{parameters1} and \eqref{parameters2}, and initial conditions as in \eqref{parameters3}. The main plot shows that a nonsingular bounce occurs, and that the time  scale of the bounce is short (it is a ``fast bounce'' model). The inner insert shows a blowup of the smooth Hubble parameters during the bounce phase: this zoomed-in view of the Hubble parameters around the bounce point shows that the Hubble rates vanish at different times, so that the scale factors bounce at different times as well.}
\label{Fig:Hubble}
\end{figure}

\begin{figure}
\includegraphics[scale=0.3]{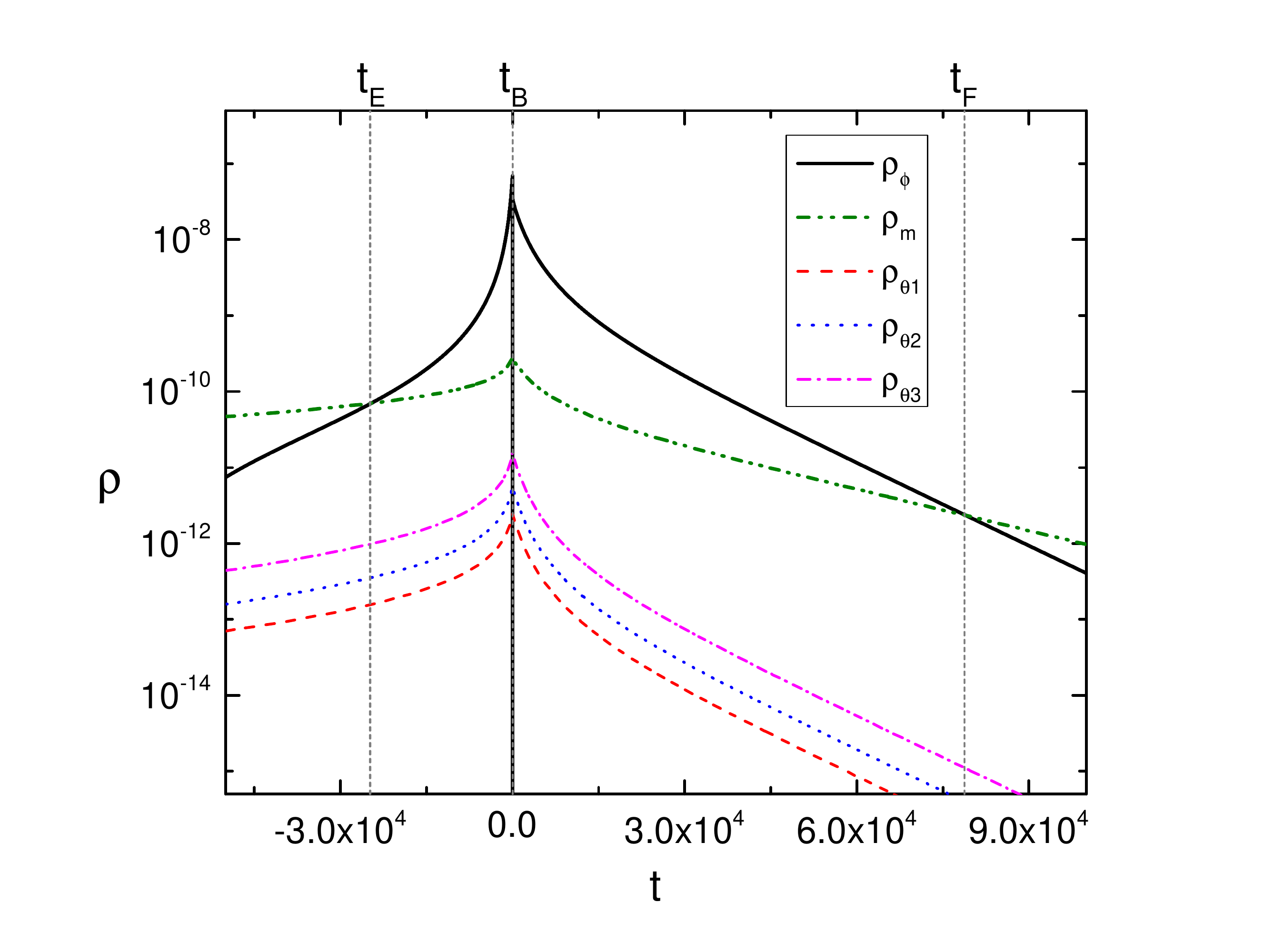}
\caption{Time evolution of the ``Effective'' energy densities of the scalar field $\rho_\phi$ (full black line), the matter fluid $\rho_\mathrm{m}$ (dot-dashed green line) and the anisotropy factors $\rho_{\theta i}$ (red dashed, blue dotted and
magenta dot-dashed lines), with same initial conditions and background parameters are in Fig. \ref{Fig:Hubble}.}
\label{Fig:rho}
\end{figure}

Moreover, we consider the following parameters of the matter fluid and the anisotropy
\begin{eqnarray}\label{parameters2}
 &\rho_{\mathrm{m},_{\mathrm{B}}} = 2.8 \times 10^{-10},~~M_{\theta,1} = 2.2 \times 10^{-6},\nonumber\\
 &M_{\theta,2} = 3.4 \times 10^{-6},~~M_{\theta,3} = -5.6 \times 10^{-6},
\end{eqnarray}
and choose as the initial conditions for the scalar field the following:
\begin{eqnarray}\label{parameters3}
 \phi_\mathrm{ini}=-2,~~\dot\phi_\mathrm{ini}=7.8\times10^{-6}.
\end{eqnarray}
The actual computation also requires the initial value of the mean Hubble parameter, which is determined by imposing the Hamiltonian constraint equation. Figs. \ref{Fig:Hubble} and \ref{Fig:rho} show the evolution of the Hubble parameters and ``effective'' energy densities for matter components and for the anisotropy, respectively.

{}From Fig. \ref{Fig:Hubble}, one can see that Hubble parameters along all spatial coordinates evolve smoothly through the bouncing point with an approximate dependence on cosmic time which is linear. The maximal value of the mean Hubble parameter, which we denote as the bounce scale $H_{_\mathrm{B}}$, is mainly determined by the value of the potential parameter $V_0$. Specifically, $H_{_\mathrm{B}}$ is of order $\mathcal{O}(10^{-4}\Mp )$ in our numerical result. We also note that the bounces occurring in the three spatial directions do not occur at exactly the same moment -- a consequence of the existence of anisotropy. This could leave a smoking gun signature for detecting nonsingular bounce cosmology in high accuracy CMB experiments since the difference in the times of the bounces along various spatial coordinates would affect the ultraviolet (UV) modes of primordial perturbations passing through the bouncing phase. We leave this issue for a forthcoming investigation.

From Fig. \ref{Fig:rho}, one can easily see that the universe in our model experiences four phases, which are matter contraction, Ekpyrotic contraction, the bounce, and fast-roll expansion, in turn. At the beginning, the universe is dominated by the matter fluid. At some point (time $t_{_\mathrm{E}}$) during the phase of contraction, the contribution of the scalar field becomes dominant, and the universe enters the Ekpyrotic phase. Note that the effective energy density of anisotropies grows faster than $\rho_\mathrm{m}$ but slower than $\rho_\phi$ during the matter contraction phase. Thus, if $\rho_{\theta i}$ does not dominate over the background before $t_{_\mathrm{E}}$, it will never become dominant throughout the whole evolution, as already discussed in the previous sections. After the bounce, the scalar field $\phi$ enters a fast roll phase with an effective equation of state equal to unity. As a consequence, the energy densities $\rho_\phi$ and $\rho_{\theta i}$ dilute at the same rate, and finally the matter fluid catches up with the density of $\phi$ at the time $t_{_\mathrm{F}}$.

\begin{figure}
\includegraphics[scale=0.25]{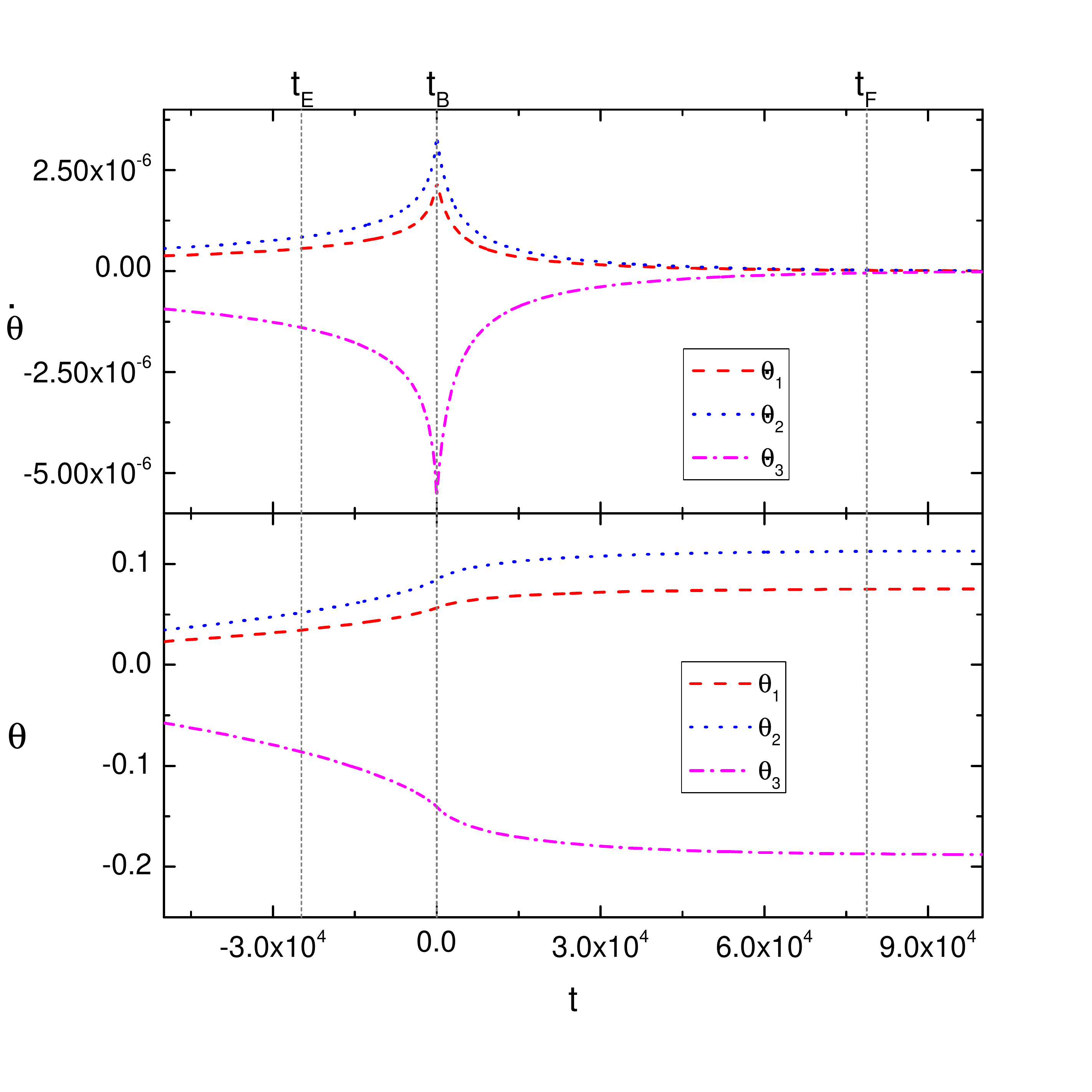}
\caption{Time evolution of the anisotropy factors $\theta_i$ (lower panel) and their time derivatives $\dot\theta_i$ (upper panel), with same initial conditions and background parameters are in Fig. \ref{Fig:Hubble}. The anisotropies increase during
the contracting phase but rapidly approach constant values in the following expanding phase.}
\label{Fig:thdth}
\end{figure}

Fig. \ref{Fig:thdth} shows the evolution of the anisotropy factors $\theta_i$ and their time derivatives. Although the anisotropy functions grow during the contraction, they evolve towards constant values in the expanding epoch. Therefore, after the time $t_{_\mathrm{F}}$, one can rescale all scale factors by absorbing the asymptotic factors in a redefinition of the coordinates, and we finally get an isotropic universe. It implies that at the level of homogeneous cosmology the anisotropies do not destabilize our nonsingular bounce model. This can also be read from the upper panel of Fig. \ref{Fig:thdth} which shows that $\dot\theta_i$ approach zero after a sufficiently long period of expansion.

\begin{figure}
\includegraphics[scale=0.25]{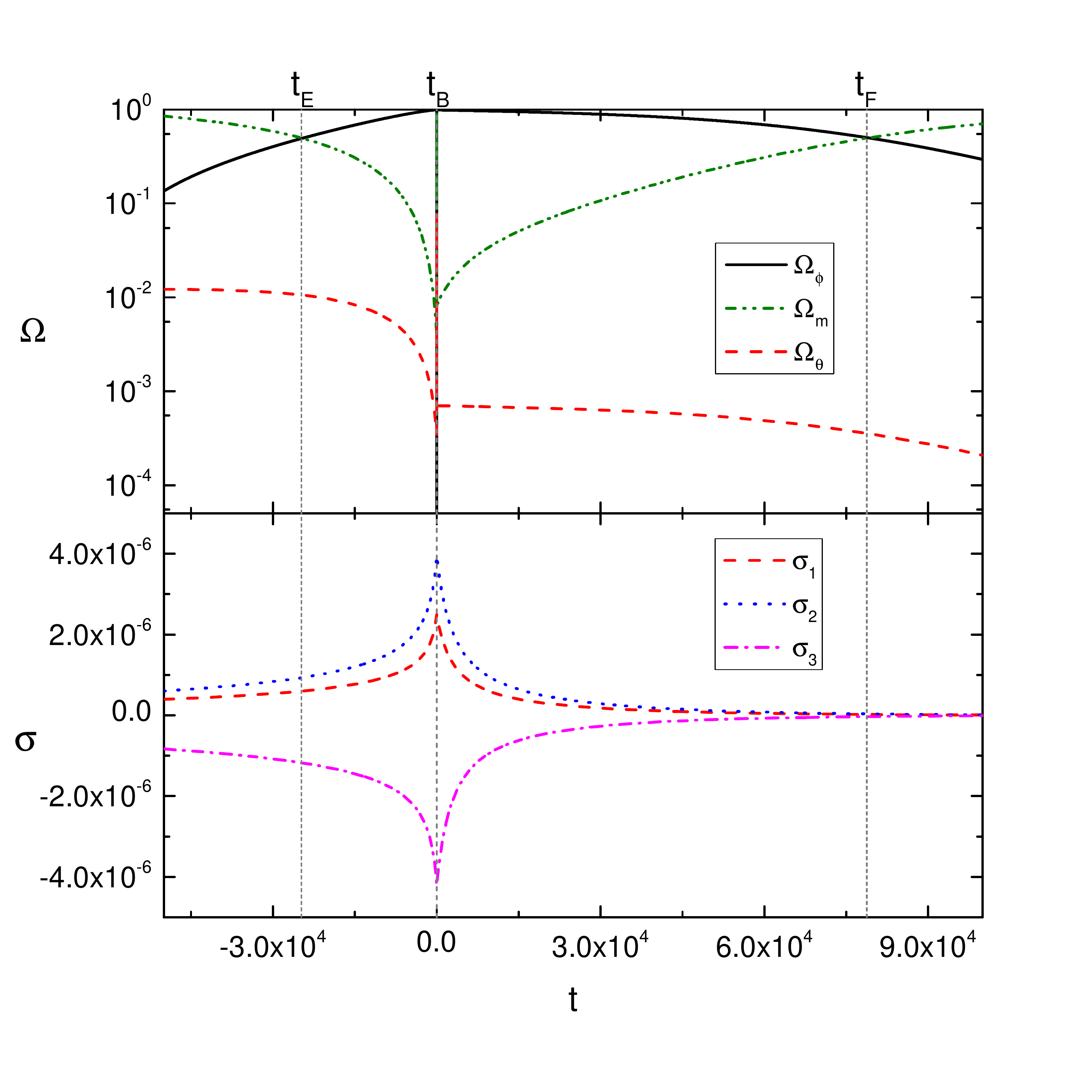}
\caption{Time evolution of the density parameters $\Omega_\phi$,
$\Omega_\mathrm{m}$, and $\Omega_\theta$ (upper panel), and
of the shear function $\sigma_i$ (lower panel), with
same initial conditions and background parameters are in
Fig. \ref{Fig:Hubble}.}
\label{Fig:siOme}
\end{figure}

In order to better characterize the anisotropy quantitatively, we can define the so-called shear parameters
\begin{eqnarray}
 \sigma_i \equiv \dot\theta_i \ex^{2\theta_i},
\end{eqnarray}
and the density parameters
\begin{eqnarray}
 \Omega_{_\mathrm{I}} \equiv \frac{\rho_{_\mathrm{I}}}{\sum_{_\mathrm{I}}\rho_{_\mathrm{I}}},
\end{eqnarray}
where the subscript ``{\scriptsize I}'' represents $\phi$, $\mathrm{m}$ and $\theta$, respectively. Fig. \ref{Fig:siOme} shows the numerical solution we obtained for their time development. The shear functions increase up to their maximal values at the bounce point, after which they rapidly decrease to end up vanishingly small when the universe connects with the Standard Big Bang evolution. From the evolution of the density parameters, we see that the contribution of the anisotropy only grows relative to the dominant density in the phase of matter contraction, but it then rapidly decreases in the Ekpyrotic phase and in the fast roll phase.

Note that all numerical calculations shown here are meant to illustrate the discussion of the previous sections. Indeed, the parameters chosen do not satisfy the bounds imposed by CMB observations, so the effects of setting non vanishing initial anisotropies can be enlarged so as to be visible at all. Assuming parameter values taking into account the experimental constraints would lead to figures exactly similar to those obtained for a regular isotropic bouncing model.

\section{Conclusion and discussion}

We have studied homogeneous but anisotropic cosmological solutions of the Galileon type action with an Ekpyrotic scalar field potential, as originally introduced in \cite{us}. The model was constructed to yield non-singular isotropic bouncing cosmological solutions. In general, the contracting phase of a bouncing cosmology suffers from a BKL instability against the growth of anisotropies. By studying the solutions of Bianchi-type homogeneous solutions in which the scale factors in different spatial dimensions are allowed to be independent we have studied the evolution of the anisotropies in our model. We have shown that the anisotropies are damped in the phase of Ekpyrotic contraction, as was already conjectured on qualitative grounds in \cite{us}. They remain small during the non-singular bounce phase, and are further suppressed once space begins to expand again. This (counter) example shows that a no-go theorem concerning smooth bounces cannot be formulated, at least in terms of BKL instabilities.

It would be of interest to study the evolution of linearized inhomogeneities in our anisotropic background and to see if there are any late time signatures of the initial (pre-Ekpyrotic) contracting phase when the anisotropies might be quite large. This is left for future work.

\begin{acknowledgments}
The work of R.B. and Y.C. is supported in part by an NSERC Discovery Grant and by the Canada Research Chair program. The research of P.P. is supported by CNRS and a PICS collaboration grant (\#05893). One of us (R.B.) thanks Paul Steinhardt and Bing-Kan Xue for discussions which stimulated us to perform this improved analysis of anisotropies in our bouncing cosmology
model.
\end{acknowledgments}


\end{document}